\documentclass[iop]{emulateapj}

\usepackage{amsmath}
\newcommand{\rsun}{R$_\Sun$~}
\newcommand{\sn}[2]{{#1} \times 10^{#2}}
\newcommand{\proc}[1]{\texttt{#1}}
\newcommand{\ionwl}[3]{\ion{#1}{#2}~{#3}~{\AA}}

\shorttitle{Modeling Nanoflares in AR~11726}
\shortauthors{Allred et al.}

\begin{document}

\title{A 3D Model of AR~11726 Heated by Nanoflares}

%\correspondingauthor{Joel C. Allred}
\email{joel.c.allred@nasa.gov}

\author{Joel C. Allred}
\affil{NASA Goddard Space Flight Center, Solar Physics Laboratory, Code 671, Greenbelt, MD 20771, USA}

\author{Adrian N. Daw}
\affil{NASA Goddard Space Flight Center, Solar Physics Laboratory, Code 671, Greenbelt, MD 20771, USA}

\author{Jeffrey W. Brosius}
\affiliation{Catholic University of America at NASA Goddard Space Flight Center, Solar Physics Laboratory, Code 671, Greenbelt, MD 20771, USA}

\begin{abstract}
The Extreme Ultraviolet Normal Incidence Spectrograph (EUNIS) and the Hinode/ EUV Imaging Spectrometer (EIS) observed AR~11726 on 2013 April 23. We present intensity images in numerous atomic lines constructed from these observations. These lines are formed over a wide range of temperatures, and we use their relative intensities to constrain a parameterization of nanoflare heating. We construct a 3D model of the magnetic field in this active region by extrapolating the surface magnetic field into the corona and using SDO/AIA images of coronal loops to ensure that extrapolated magnetic field lines co-align with observed coronal loops. We trace 2848 magnetic field lines within the volume of this active region and model how they fill with hot plasma in response to nanoflare heating. We perform a parameter study to determine how the frequency and energy released in nanoflares scale with magnetic field strength and loop length. From our 3D model, we construct synthetic images of the lines observed by EUNIS and EIS and constrain the parameter study by minimizing the difference between the synthetic and observed images. 
%We find that in the best fit model, the nanoflare frequency is a power-law that scales inversely with energy, $\alpha = -2.4$, magnetic field strength, $\beta = 1.5$, and inversely with loop length, $\gamma = -1.0$.
\end{abstract}


\section{Introduction} \label{sec:intro}
Explaining the mechanisms that maintain hot plasma ($>1$~MK) in solar corona active regions has been a longstanding unresolved question in solar astrophysics. Many previous studies have identified the coronal magnetic field as the primary energy source powering this heating \citep[e.g., see the review in][]{2006SoPh..234...41K}, and images of coronal active regions have shown that they are highly structured into loops which are thought to trace out the magnetic field. It is likely within these loops that magnetic energy is converted into heat. One broad group of theories describing this process involves the dissipation of Alfv\'en waves \citep{1978ApJ...226..650I,1984ApJ...277..392H, 1987ApJ...317..514D, 2011ApJ...736....3V, 2011Natur.475..477M} and another involves impulsive heating at small scales likely due to reconnection of filamented loops at sub-resolution (i.e., nanoflares) \citep{1975ApJ...199L..53G, 1984ApJ...283..421L, 1988ApJ...330..474P, 1994ApJ...422..381C}. 

In the nanoflare scenario, independent heating events occurring on individual strands within coronal loops can impulsively raise the strand plasma to temperatures $\sim$~6~-~10~MK, which is much hotter than the average active region temperature of 2~-~4~MK. The heating time is quite fast perhaps $<100$~s, but the cooling time through thermal conduction and radiation can be much slower. Thermal conduction also drives chromospheric ablation \citep[historically referred to as evaporation;][]{antonucci1999, bornmann1999, 2011SSRv..159...19F} which populates the loop with denser plasma. The net effect of many such events occurring at random throughout the corona results in the observed ``average'' corona. 

The nanoflare model has recently attracted more attention because of new observations that demonstrate the existence of very hot plasma in closed loops within an active region. \citet{2017NatAs...1..771I} detected faint but very hot ($> 10$ MK) plasma using the Focusing Optics X-ray Solar Imager (FOXSI-2) sounding rocket. \citet{2014ApJ...790..112B} report on an observation from the Extreme Ultraviolet Normal Incidence Spectrograph (EUNIS-13) of AR~11726. They observed faint but pervasive emission in the atomic line, \ionwl{Fe}{19}{592.2}, which forms in the temperature range 6~-~11~MK. Such hot plasma may be an indication of nanoflare heating. \citet{2012ApJ...753...35V} have detected the signature of plasma cooling from a peak temperature of at least 7~MK by looking at time delays between filters on the Atmospheric Imaging Assembly (AIA). \citet{2014Sci...346B.315T} and \citet{2018ApJ...856..178P} showed that Doppler shifts observed in transition region bright points is most consistent with impulsive heating low in the coronal loop as predicted from nanoflare simulations.

\citet{1991SoPh..133..357H} noted that the flare frequency is related to the energy released by a power-law, with index, $\alpha$. \citet{1991ApJ...380L..89L} showed this power-law distribution can arise when the coronal magnetic field is in a self-organized critical state. For scales of large flares down to microflares, it is relatively straightforward to determine $\alpha$ directly, and perhaps a similar index applies even to the scale of nanoflares, but this is not certain. Many authors have attempted to constrain this index. For example, \citet{2000ApJ...529..554P} performed a statistical study of events observed in the EUV filters on Transition Region and Coronal Explorer (TRACE). Following up that work, \citet{2002ApJ...572.1048A} used observations from EUV filters on TRACE combined with soft X-rays from Yohkoh Soft X-ray Telescope (SXT) and obtained a value for $\alpha$ that is consistent with the self-organized critical theory. A more complete discussion of this subject is presented in \citet{2006SoPh..234...41K}.

The EUNIS-13 observation of AR~11726, first described in \citet{2014ApJ...790..112B}, was coordinated to have co-temporal observations from the Hinode/ EUV Imaging Spectrometer \citep[EIS;][]{2007SoPh..243....3K, 2007SoPh..243...19C}, and of course, continuous monitoring from the instruments aboard Solar Dynamics Observatory \citep[SDO;][]{2012SoPh..275....3P}. This wealth of data provides a unique opportunity to constrain the emission measure in that active region over a wide range of temperatures. For this work, we parameterize the nanoflare frequency distribution as a power-law and use these data to constrain the slope and scale of that distribution. We extrapolate the observed surface magnetic field into the corona and trace thousands of closed field lines within the active region volume. We interpret these field lines as the centers of coronal loops and perform simulations of nanoflare heating within them by randomly sampling the parameterized nanoflare frequency distribution. We use the resulting temperature and density stratification within the loops--- Sun: magnetic fields to predict atomic line emission. These are projected along the Earth's line of sight, and we compare these synthetic images with the EUNIS and EIS observations. By varying parameters and minimizing the difference between prediction and observation, we constrain the nanoflare frequency distribution.

This paper is organized as follows. In \S\ref{sec:obs} we describe EUNIS, EIS and SDO coordinated observations of AR~11726. In \S\ref{sec:3dmodel}, we demonstrate how we have constructed a 3D model of the magnetic field in this active region, and in \S\ref{sec:nanoflares} we describe using hydrodynamic simulations of nanoflares to model populating these loops with hot plasma and the resulting emission. We present the results of a parameter study designed to constrain the frequency distribution of nanoflares in \S\ref{sec:parameterstudy}, and finally in \S\ref{sec:conc} we discuss these results and draw conclusions. 

\section{SDO, EUNIS, and EIS Observations of AR~11726}\label{sec:obs}
\subsection{SDO\label{sec:sdo}}
We used the six coronal channels of the SDO/AIA instrument \citep{2012SoPh..275...41B,2012SoPh..275...17L} (i.e., AIA 94, 131, 171, 193, 211, and 335) to provide contextual alignment for EUNIS and EIS and for loop tracing (\S\ref{sec:3dmodel}). The times of the AIA images were chosen to closely match the EUNIS observation (2013-04-23 17:32:00 UT). AIA level-1 FITS files observed at this time were downloaded using the Virtual Solar Observatory (VSO) service (SSW IDL routine, \proc{VSO\_SEARCH}). The filter images and headers (including pointing information) were read using \proc{READ\_SDO}. Similarly, the SDO/Helioseismic and Magnetic Imager \citep[HMI;][]{2012SoPh..275..207S,2012SoPh..275..229S} line-of-sight magnetogram was obtained and read using the \proc{VSO\_SEARCH} and \proc{READ\_SDO} procedures.

\subsection{EUNIS}
EUNIS is a two-channel imaging spectrometer designed to observe the Sun with high temporal and spatial resolutions. Spectra are taken in 692 pixels along slits with length 636\arcsec. Thus, each pixel covers 0.\arcsec92, which oversamples the spatial resolution of the optical system ($\sim3$\arcsec) by about a factor of 3. To reduce noise and ensure that the effective spatial resolution matches the optical system, three neighboring pixels are averaged together. The slits move across the field of view at a rate of 2.\arcsec14~s$^{-1}$ in a direction perpendicular to the slit's length, and taking 1.32~s exposures. This yields an effective spatial resolution of 2.\arcsec77 $\times$ 2.\arcsec82. The two channels cover the wavelength ranges of 525 - 635 {\AA} and 300 - 370 {\AA} and include lines that form at temperatures ranging from 0.03 - 10 MK, but for this study we have focused on lines in the temperature range 1 - 10 MK. These lines are listed in Table~\ref{tab:linelist}. 

EUNIS was flown on a sounding rocket on 23 April 2013 and observed AR~11726. For this work, we use 32 exposures that comprise the first full northward-directed raster of AR~11726 (17:32:45- 17:33:26 UT). The central time of this 41 s raster is 17:33:06 UT. Line intensities and uncertainties were measured by fitting the line profiles to Gaussians as discussed in \citet{2014ApJ...790..112B}. The low and high channel observations were co-aligned to the AIA 193 channel using the \ionwl{Fe}{12}{338.3} and \ionwl{Fe}{12}{592.6} images, respectively. In this procedure, we varied the Solar-X and -Y coordinates of EUNIS's slit center, rotation angle, and pixel sizes to minimize the $\chi^2$ differences between these and the AIA 193 image. The slits were rotated by an angle of $3.3^{\circ}$ clockwise relative to the solar east-west axis. The images were centered at solar positions, (659.\arcsec6, 251.\arcsec5) and (660.\arcsec7, 251.\arcsec5), for the low and high channel, respectively. More instrument details and the observing sequence are discussed in \citet{2014ApJ...790..112B}.  

%EUNIS count rates were calibrated to absolute intensity units using the following procedure. ***ADRIAN. WE NEED CALIBRATION TEXT HERE ***

\subsection{EIS}
Hinode/EIS performed coordinated observations of AR~11726. EIS is a two-channel imaging spectrometer covering the ranges 170-210 {\AA} and 250-290 {\AA}. Numerous atomic lines formed over a wide range of temperature are included in these bands. The lines observed by EIS and used in this study are listed in Table~\ref{tab:linelist}. Level-0 data stored in FITS files were converted into Level-1 using the SSW IDL routine, \proc{EIS\_PREP}. Similar to \citet{2012ApJ...744...14Y}, we found that in addition to the standard wavelength correction computed for each pixel, an additional wavelength correction of $\sim 0.02$\ {\AA} was needed to be added to the long wavelength channel. This correction was obtained from the assumption that the Doppler shifts in the \ionwl{Fe}{13}{252.0} and \ionwl{Fe}{13}{202.0} are identical. Line intensities and corresponding uncertainties were obtained using Gaussian fits to the line profiles and were performed using the routine, \proc{EIS\_AUTO\_FIT}, and using the calibration correction of \citet{2013A&A...555A..47D}. Bad pixels were removed during the line fitting procedure.

\begin{figure*}
\epsscale{.95}
\plotone{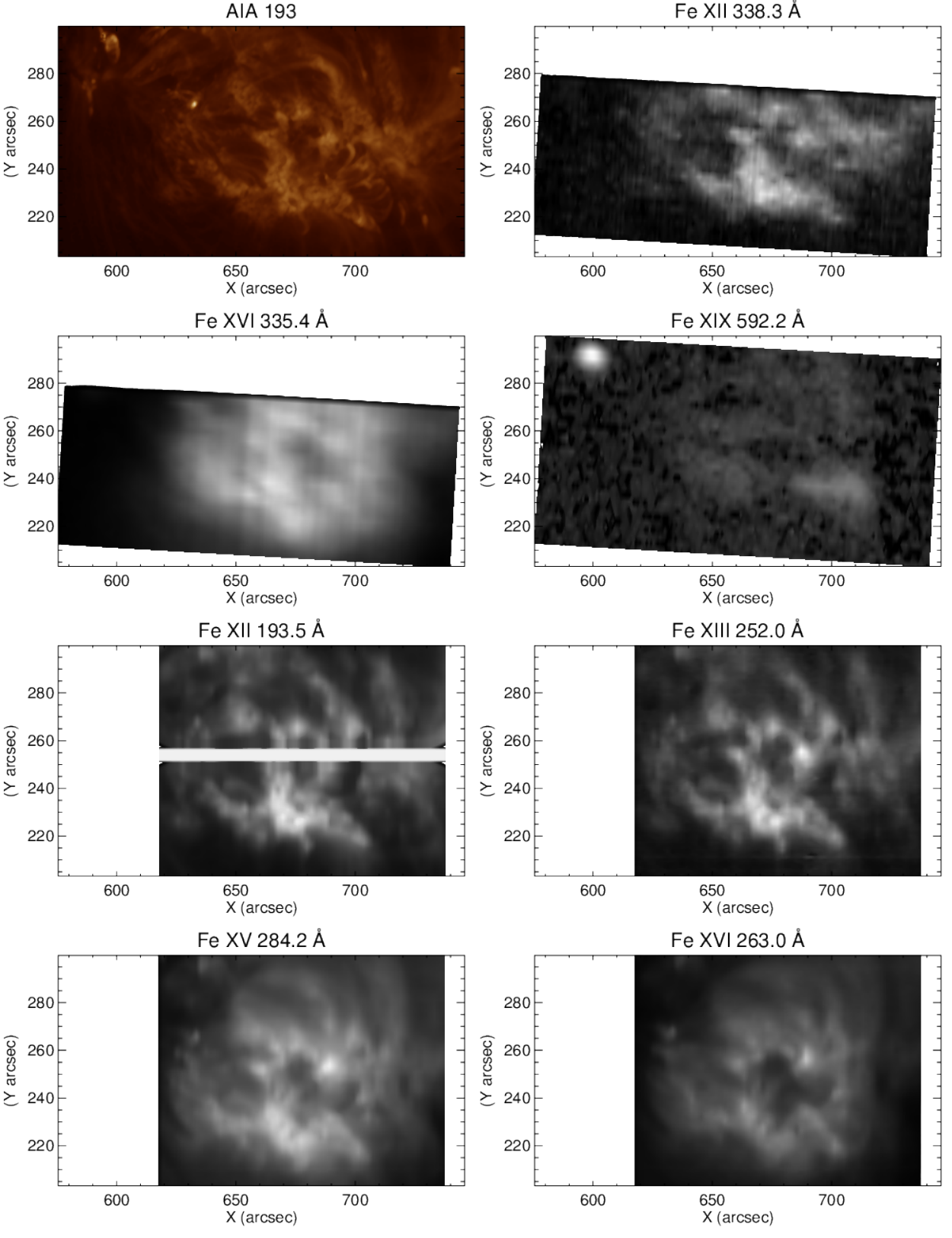}
\caption{The top left panel shows an AIA 193 image of AR~11726 taken at 2013-04-23 17:32:07.84 UT. The subsequent panels show intensity map images for several lines observed by EUNIS and EIS.  The white band near the center of the \ionwl{Fe}{12}{193.5} image indicates a region where line profile fits could not be performed because of missing data.\label{fig:lineimages}}
\end{figure*}

We co-aligned the EIS low and high channel observations to the AIA 193 channel using the \ionwl{Fe}{12}{193.5} and \ionwl{Fe}{13}{252.0} images and implementing a $\chi^2$ minimization procedure similar to what was described above for EUNIS. The EIS slits have a width of 2\arcsec and were oriented along the north-south axis and moved along the east-west axis taking a total of 61 50~s exposures. The effective pixel size is 2.\arcsec00 $\times$ 1.\arcsec00 with a resolution of 3~-4~\arcsec. The images were centered at solar positions, (677.\arcsec7, 243.\arcsec5) and (677.\arcsec3, 259.\arcsec7), for the low and high channels, respectively. Images of several of the EIS and EUNIS line intensities together with the AIA 193 channel are shown in Figure~\ref{fig:lineimages}.

\begin{deluxetable}{lccc}
  \tablecaption{List of lines used in this study.\label{tab:linelist}}
  \tablecolumns{4}
  \tablewidth{0pt}
  \tablehead{
    \colhead{Ion} &
    \colhead{Wavelength ({\AA})} &
    \colhead{log T$_{max}$ (K)} &
    \colhead{Observing Instr.} 
  } 
  \startdata
  \ion{Fe}{10}   & 184.5 & 6.1 & EIS   \\
\ion{Fe}{10}   & 190.0 & 6.1 & EIS   \\
\ion{Al}{10}   & 332.8 & 6.1 & EUNIS \\
\ion{Si}{10}   & 258.4 & 6.2 & EIS   \\
\ion{Fe}{11}   & 202.4 & 6.2 & EIS   \\
\ion{Fe}{11}   & 188.3 & 6.2 & EIS   \\
\ion{Fe}{11}   & 180.4 & 6.2 & EIS   \\
\ion{Fe}{11}   & 182.2 & 6.2 & EIS   \\
\ion{Fe}{11}   & 181.1 & 6.2 & EIS   \\
\ion{Fe}{12}   & 338.3 & 6.2 & EUNIS \\
\ion{Al}{11}   & 550.0 & 6.2 & EUNIS \\
\ion{Al}{11}   & 568.1 & 6.2 & EUNIS \\
\ion{Si}{11}   & 580.9 & 6.2 & EUNIS \\
\ion{Fe}{12}   & 592.6 & 6.2 & EUNIS \\
\ion{Fe}{12}   & 195.1 & 6.2 & EIS   \\
\ion{Fe}{12}   & 193.5 & 6.2 & EIS   \\
\ion{Fe}{12}   & 192.4 & 6.2 & EIS   \\
\ion{Fe}{13}   & 252.0 & 6.2 & EIS   \\
\ion{Fe}{13}   & 202.0 & 6.2 & EIS   \\
\ion{Fe}{13}   & 197.4 & 6.2 & EIS   \\
\ion{Fe}{13}   & 201.1 & 6.2 & EIS   \\
\ion{Fe}{13}   & 204.9 & 6.2 & EIS   \\
\ion{Fe}{13}   & 200.0 & 6.2 & EIS   \\
\ion{Fe}{14}   & 334.2 & 6.3 & EUNIS \\
\ion{Fe}{14}   & 274.2 & 6.3 & EIS   \\
\ion{Fe}{14}   & 270.5 & 6.3 & EIS   \\
\ion{Fe}{14}   & 252.2 & 6.3 & EIS   \\
\ion{Fe}{14}   & 264.8 & 6.3 & EIS   \\
\ion{Fe}{15}   & 327.0 & 6.3 & EUNIS \\
\ion{Fe}{15}   & 284.2 & 6.3 & EIS   \\
\ion{S}{13}    & 256.7 & 6.4 & EIS   \\
\ion{Fe}{16}   & 335.4 & 6.4 & EUNIS \\
\ion{Fe}{16}   & 251.1 & 6.4 & EIS   \\
\ion{Fe}{16}   & 263.0 & 6.4 & EIS   \\
\ion{Ca}{14}   & 193.9 & 6.6 & EIS   \\
\ion{Fe}{19}   & 592.2 & 6.9 & EUNIS \\
\ion{Fe}{20}   & 567.9 & 7.0 & EUNIS \\

  \enddata
\end{deluxetable}

\section{Constructing a 3D Model of AR~11726} \label{sec:3dmodel}
\begin{figure*}[htp]
\plottwo{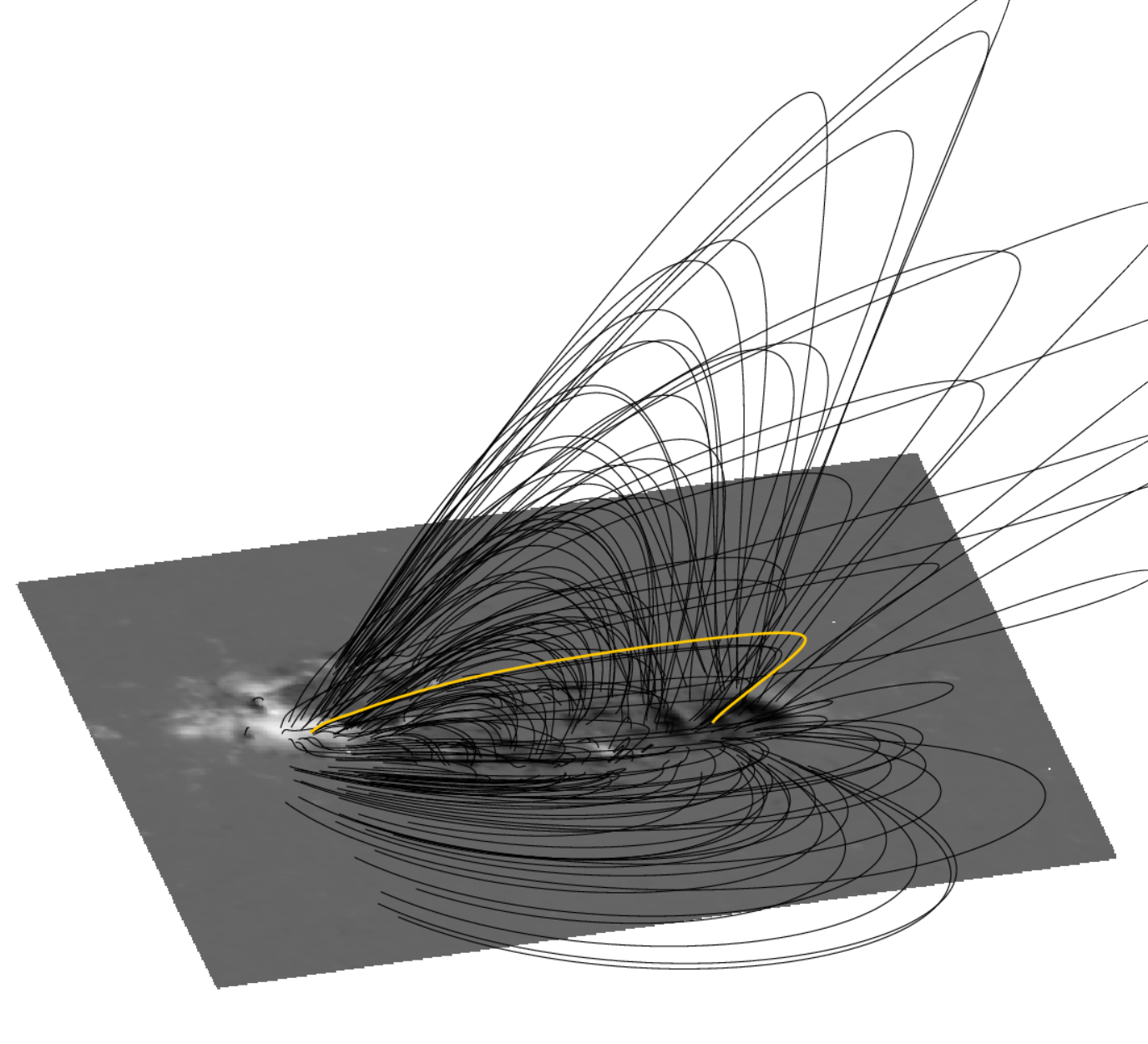}{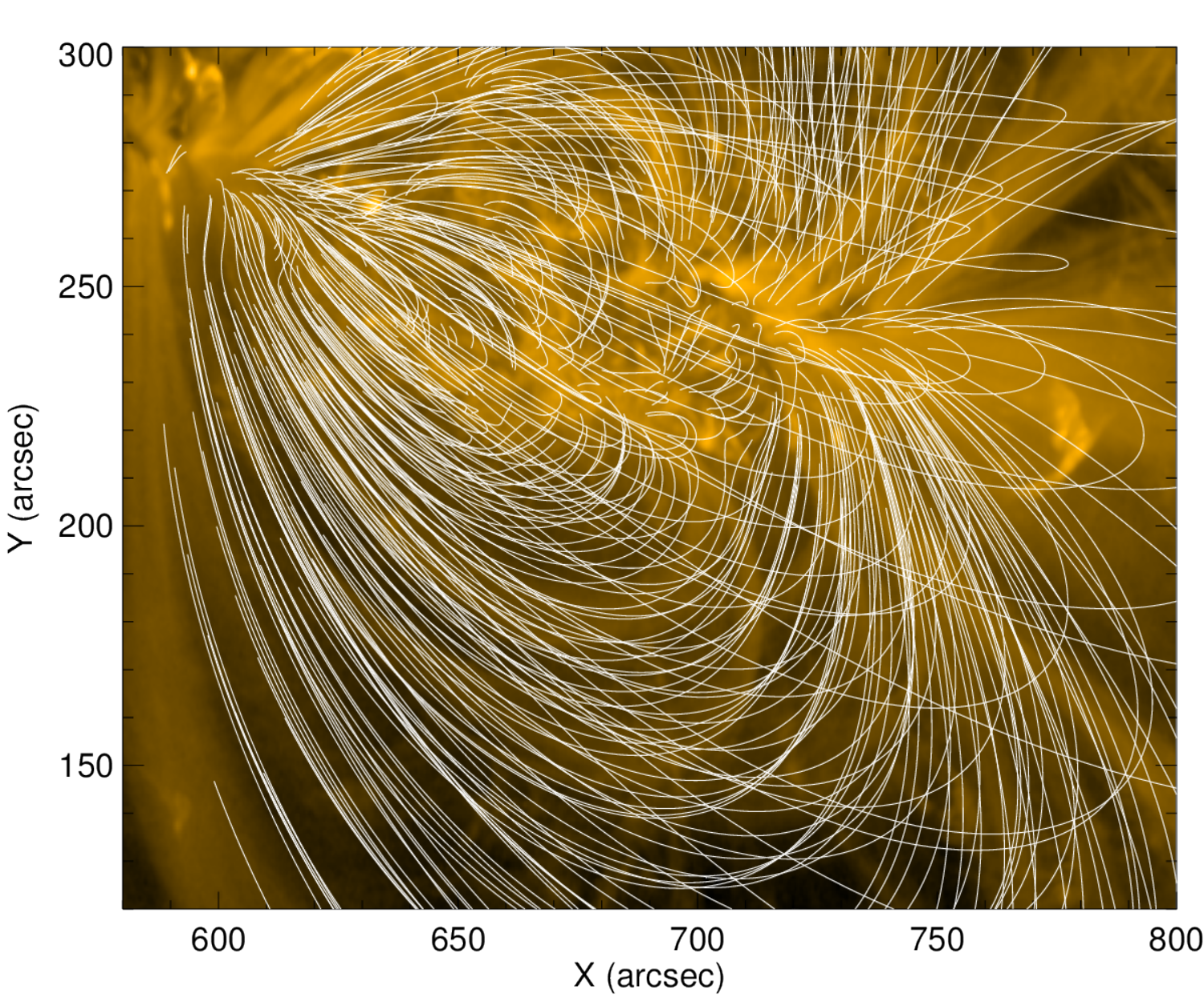}
\caption{(left) The positions of a representative set of field lines shown superimposed over an HMI LOS magnetogram. One field line is highlighted in yellow to indicate the position of the example loop described in \S\ref{sec:1dmodel}. (right) The same field lines superimposed on an AIA 171 image. \label{fig:loops}}
\end{figure*}

We have constructed a 3D model of AR~11726 by extrapolating the photospheric magnetic field observed by SDO/HMI into the corona using the Vertical Current Approximation Non-linear Force Free Field (VCA-NLFFF) technique described in \citet{2013ApJ...763..115A,2016ApJS..224...25A}. The key advantage of this technique is its ability to identify loop structures in AIA images and minimize the misalignment of predicted magnetic field lines compared with the position of observed loops. This is important for comparisons of our AR model to observed images discussed in \S~\ref{sec:compare}. For this study, we have configured the VCA-NLFFF code to use 150 potential and non-potential sources and considered a 0.5~R$_\Sun$ field of view centered at the heliographic position, N11W47.  AIA and HMI images were selected to most closely match the time of the EUNIS observation (\S\ref{sec:sdo}). Only the coronal AIA channels were used by the VCA-NLFFF code, i.e., the AIA 94, 131, 171, 193, 211, and 335 channels. With these parameters the VCA-NLFFF code calculated the 3D magnetic field within a volume centered around N11W47 and extending 0.5~\rsun in the heliocentric Cartesian $x$ and $y$ directions and to a height of 1.5~\rsun in the $z$ direction. These are calculated on a regular heliocentric Cartesian grid with resolution of $315 \times 315 \times 430$. Within this volume, we have traced magnetic field lines in each voxel. Those field lines that close within the volume and have a minimum loop length of 1 Mm become part of our ensemble. The minimum loop length criterion was chosen to eliminate very short loops that would not be observable by current instruments. We found 2848 field lines fit these criteria. We interpret these field lines as the centers of closed loops with circular cross sections. The cross sectional area of the loops is assumed to expand into the corona so that magnetic flux is conserved within the loop. Formally, the cross sectional area, $A(s)$, is derived from the equality, $A(s) B(s) = A(s = 0) B(s = 0) $, where $s$ is the distance along the loop and $B$ is the magnetic flux density. The radius of the loop at $s = 0$ is assumed to be 220 km (0\arcsec.3) for each loop which is comparable to loop radii observed by high-resolution imagers such as Hi-C \citep{2014SoPh..289.4393K} and comparable to the average loop width determined by \citet{2017ApJ...840....4A}. The mean loop radius including the expansion factor is 620 km. Figure~\ref{fig:loops} shows the 3D position of a representative set of field lines and their projection onto an AIA 171 image. 

Since the area expansion of each loop is calculated independently, it is possible for the cross sectional areas of neighboring loops to overlap. When calculating emission from these loops, it important that an overlapping volume is not double-counted. When a voxel has more than one loop going through it, the emission originating from that voxel is determined by averaging the emission from each of the corresponding loops. 

\section{Simulating the Hydrodynamic Response to Nanoflare Heating} \label{sec:nanoflares}
\subsection{Nanoflare Frequency Distribution}
For this work, we want to explore how the frequency of nanoflare heating events scales with the energy released in the events, the local magnetic flux density and the length of the magnetic flux loops where they occur. To that end we parameterize the frequency distribution function of nanoflare heating and study how varying parameters best reproduce the coronal emission in atomic lines formed over a broad temperature range observed by EUNIS and EIS. We assume that the frequency, $f$, of occurrence of a nanoflare with energy between $E$ and $E + \textrm{d}E$ in the volume, $\textrm{d}V$, is given by

\begin{equation}
 f = f_0 (E/E_0)^\alpha (B/B_0)^\beta (L/L_0)^\gamma \textrm{d}E \textrm{d}V, 
 \label{eq:freq}
\end{equation}
where $E_0 = 10^{23}$~erg, $B_0 = 5$~G, and $L_0 = 50$~Mm are characteristic scale sizes of the nanoflare energy, magnetic flux density, and loop length, respectively. $\alpha$, $\beta$, and $\gamma$ are power-law index parameters that determine how the frequency scales with energy, magnetic flux density and loop length, respectively. We define $f_0$ in terms of a parameter, $q_0$, by 

\begin{equation}
  f_0 \equiv
  \begin{cases}
       \frac{q_0(\alpha - 2)}{E_0^2 \left[1 - (E_0/E_f)^{\alpha-2}\right]} & \text{if $\alpha \neq 2$} \\
       \frac{q_0}{E_0^2 \ln{\frac{E_f}{E_0}}}                                       & \text{if $\alpha = 2$} \\
  \end{cases}
  \label{eq:f0}
\end{equation}
$q_0$ is related to the time-averaged volumetric heating rate. That heating rate is given by

\begin{equation}
 \int_{E_0}^{E_f} f E \textrm{d}E = q_0 (B/B_0)^\beta (L/L_0)^\gamma
 \label{eq:hr}
\end{equation} 
$E_0$ and $E_f$ are the smallest and largest energy events considered in this nanoflare frequency distribution. We assume $E_f = 10^{27}$ erg, which is comparable to the energy released in a small microflare. 

In summary, the free parameters characterizing the frequency distribution are $\alpha$, $\beta$, $\gamma$, and $q_0$. In theory, the duration of a nanoflare heating event is also a parameter. However, we have found that the time-averaged differential emission measure (DEM) only weakly depends on the duration. Dissipating the nanoflare energy faster produces a hotter transient, but the heat flux quickly moves that energy throughout the loop resulting in a similar temperature structure to what would be obtained had the energy been dissipated more slowly. %This is illustrated in Figure~\ref{fig:comparedur}, which compares spatially integrated DEMs obtained using identical nanoflare frequency distributions but with different nanoflare durations. The black and red lines represents the DEMs for cases in which the nanoflare durations are assumed to be constant at 100~s and 10~s, respectively. These cases are nearly identical below 6~MK, and therefore will predict very similar line emission in all but the hottest lines. The 10~s case has more emission measure at higher temperatures, but in both cases the hot emission measure is several orders of magnitude below the peak around 2~MK. 
For the study described in \S\ref{sec:parameterstudy}, we have chosen a uniform duration for all nanoflare events of 100~s. 
%\begin{figure}
% \plotone{comparedur.eps}
% \caption{A comparison of the spatially integrated (over the entire computational domain) DEMs for identical nanoflare frequency distribution functions (Eq.~\ref{eq:freq}) but with varying nanoflare duration times. Constant nanoflare durations of 100~s and 10~s are represented by the black and red lines, respectively.\label{fig:comparedur}} 
%\end{figure}

\subsection{1D Hydrodynamic Loop Models\label{sec:1dmodel}}
In active regions, the magnetic pressure is much greater than the gas pressure meaning that the magnetic force confines charged particles to move along field lines. The dynamics then are well-represented as occurring within one-dimensional loop structures (albeit with spatially varying cross-sectional areas). In this work, we perform simulations of nanoflares occurring in each of the 2848 loops traced in \S\ref{sec:3dmodel}. The hydrodynamic response of each loop is assumed to be independent of other loops. 

These simulations are performed using the ARC7 code \citep{allred2015}. ARC7 was designed to solve the MHD equations in 2.5D using a finite volume flux-corrected transport scheme \citep{1973JCoPh..11...38B}, but we have removed the magnetic field equations and reduced the code to solving hydrodynamics in a 1D geometry, with that dimension being the axis of a magnetic loop. As mentioned in \S\ref{sec:3dmodel}, the cross sectional area of a loop expands into the corona in order to conserve magnetic flux. ARC7 includes this expansion of area in calculating conservation of mass, momentum, and energy fluxes.  Both boundaries of the loop are held at a temperature of 20~kK. The density of the boundary is typically about $\sn{1}{13}$~cm$^{-3}$ but that and the boundary velocity are allowed to vary in order to maintain a non-reflecting boundary condition. 

ARC7 solves conservation equations for mass, momentum and energy on a grid with resolution, $\textrm{d}s = 385$~km. A typical loop length is $\sn{1}{5}$~km, resulting in about 260 grid cells. We have performed simulations at higher resolutions and found negligible differences, so we assume that this resolution is sufficient. The momentum equation includes gravitational acceleration, $g = g_0 (R_\Sun/r)^2 (\textbf{ds} \cdot \textbf{\^{r}})$, where $g_0$ is the acceleration at the solar surface, $r$ is the radial distance from solar center, and $\textbf{\textrm{d}s} \cdot \textbf{\^{r}}$ is the component of the loop axis along the vertical direction. The energy equation includes the effects of Spitzer thermal conduction and optically-thin radiative losses. The radiative loss function is tabulated using the CHIANTI \citep[v.~8.0.7;][]{1997A&AS..125..149D,2015A&A...582A..56D} routine, \proc{RAD\_LOSS}. In order to generate stable starting loop conditions, we add a small heating term and allow the loops to relax to a state of hydrostatic equilibrium. The heating term maintains the coronal temperature at approximately 0.1~MK and the coronal density at $\sim10^7$~cm$^{-3}$. These very low values were chosen to ensure that any emission measure above 1~MK is not affected by this background heating. The equilibrium solution is used as a starting state for the simulations presented below.

To simulate nanoflares we add an additional heating term to the energy equation. The probability of a nanoflare with energy, $E$, occurring within a computational grid cell with volume, \textrm{d}V, is given by the frequency distribution function (Eq.~\ref{eq:freq}) using the local magnetic flux density and the overall loop length. During a simulation ARC7 randomly samples the distribution function for each grid cell to determine if a nanoflare occurs there. If so, the nanoflare heating term dissipates the energy, $E$, over 100~s by linearly increasing the heating for 50~s and then linearly decreasing back to zero for 50~s. We simulate the evolution of the loops for $t_f = \sn{1}{5}$~s. This is enough time so that even relatively infrequent events will have a chance of occurring. 

\begin{figure*}
 \plottwo{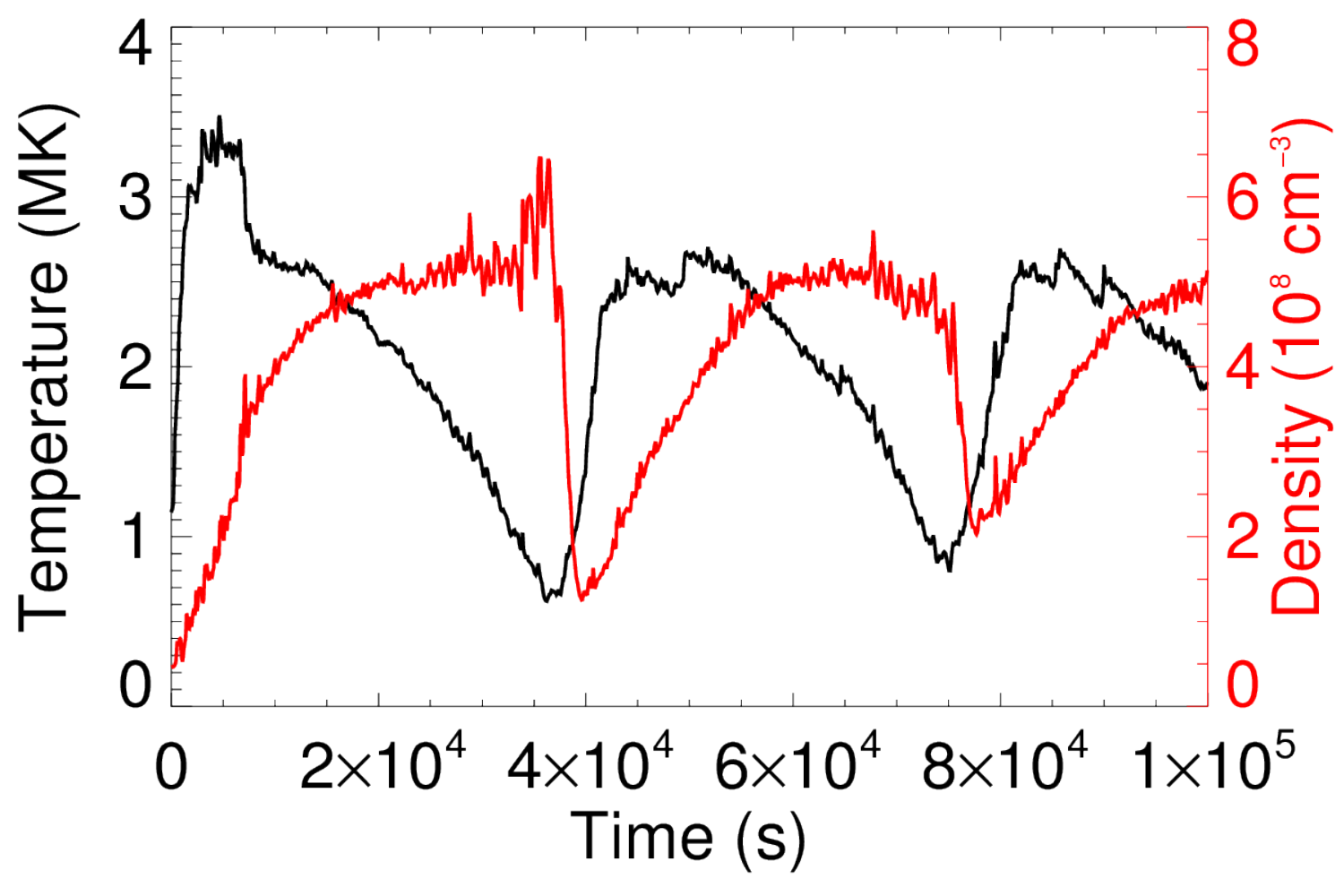}{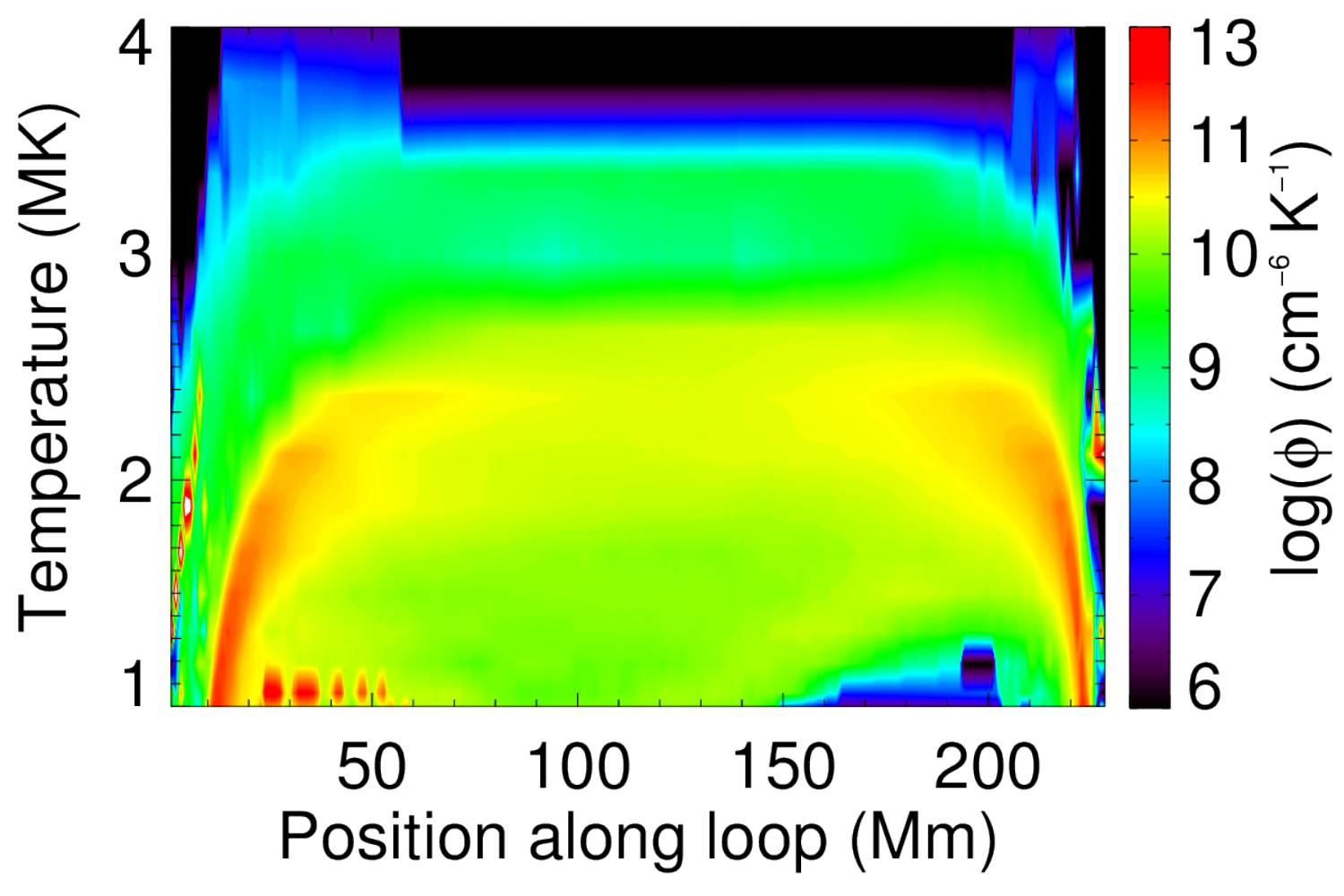}
 \caption{(left) The evolution in time of the plasma temperature (black) and electron density (red) at the loop apex of the example loop highlighted in yellow (Fig.~\ref{fig:loops}) in response to randomly generated nanoflare heating. For this simulation the frequency distribution parameters are $\alpha = -2.5$, $\beta = 1.5$, $\gamma = -1.5$, and $q_0 = \sn{5}{-4}$~erg~cm$^{-3}$~s$^{-1}$. (right) The resulting DEM field, $\phi$, as a function of position along the loop.\label{fig:loopevol}}
\end{figure*}

As an example, we show in Figure~\ref{fig:loopevol} the evolution of the loop highlighted in yellow in Figure~\ref{fig:loops} for a particular set of parameters. In this case, $\alpha = -2.5$, $\beta = 1.5$, $\gamma = -1.5$, and $q_0 = \sn{5}{-4}$~erg~cm$^{-3}$~s$^{-1}$. The left panel shows the temperature and density at the loop apex. A nanoflare early in the simulation quickly causes the apex temperature to rise to 3 MK. This is followed by a much smaller event which increases the temperature to 3.5 MK. The apex cools until about $t = \sn{3.7}{4}$~s when it is hit again by another nanoflare and quickly heats again. After the first event, the apex density gradually rises as chromospheric material ablates into the corona. When another event happens the density quickly drops since the sudden increase in pressure forces material away from the explosion, but then gradually increases again from the chromospheric ablation. The right panel shows the time-averaged DEM field, $\phi$ (Eq.~\ref{eq:demf}; discussed below in \S\ref{sec:3demission}), as a function of position along the axis of the loop. The legs of the loop are cooler than the apex and dominated by temperatures in the range 1 - 2 MK, although there is a small emission measure up to 4 MK at the cusp of the legs at positions 30 - 50 Mm and 200 - 210 Mm. 

\subsection{Constructing Emission in the 3D Volume} \label{sec:3demission}
The simulated evolution of each loop is independent of the others, i.e., there is no reason to believe that $t = 0$ for one loop corresponds to the same absolute time as $t = 0$ for some independent loop, so our method does not construct a time-dependent 3D model. Rather, from the time evolution of each loop, we wish to construct a 3D representation of the time-averaged emission resulting from all loops.

The time-averaged radiated power per unit volume from an atomic line originating at any point in space, $\textbf{x}$, is obtained from the plasma density and the line contribution function, $G(N_e, T)$,  using the following formula: 

\begin{equation}
 I(\textbf{x}) = \frac{Ab}{t_f} \int\limits_{0}\limits^{t_f}{G\left(N_e(\textbf{x},t),T(\textbf{x},t)\right) N_e(\textbf{x},t) N_H(\textbf{x},t) \textrm{d}t},
\end{equation}
where $t + \textrm{d}t$ is the period of time over which plasma at the position, \textbf{x}, has a temperature, $T$, and electron density, $N_e$. The hydrogen density, $N_H$ is assumed to be related to the electron density by $N_H = 0.83 N_e$. The abundance, $Ab$, is from the tables in \citet{2012ApJ...755...33S}. The contribution functions, $G(N_e,T)$, are obtained from the CHIANTI database and its corresponding ionization equilibrium table. We assume that the lines studied here are optically-thin so that the intensity incident on an observer is simply an integral of $I(\textbf{x})$ over the observer's line of sight. 

Similarly, we define a time-averaged differential emission measure field, $\phi(\textbf{x},T)$, by 

\begin{equation}
 \phi(\textbf{x},T) \equiv \frac{1}{t_f \textrm{d}T} \sum\limits_{t_i}{n_e(\textbf{x},t_i) n_H(\textbf{x},t_i) \textrm{d}t},
 \label{eq:demf}
\end{equation}
where $t_i + \textrm{d}t$ is a time interval over which the position, $\textbf{x}$, has temperature in the interval $T + \textrm{d}T$. The DEM experienced by an observer is obtained by integrating $\phi$ along the observer's line of sight,

%\begin{equation}
%  \textrm{DEM}(T) = \int \phi \textrm{d}l 
%\end{equation}

\section{Parameter Study} \label{sec:parameterstudy}
The frequency of nanoflare heating is parameterized using Eqs.~\ref{eq:freq}~-~\ref{eq:f0}. We constrain the parameters, $q_0$, $\alpha$, $\beta$, and $\gamma$, by varying them and comparing the resulting predicted emission with the EUNIS and EIS observations. The ranges over which these parameters were varied is shown in Table~\ref{tab:paramrange}. First, these were varied using a resolution listed in the ``Course Step'' column. Once best values were found at this resolution, we used the resolutions listed in the ``Fine Step'' column to fine tune the parameter study. For each set of parameters, the hydrodynamic response to nanoflare heating was modeled using the procedure described in Section~\ref{sec:nanoflares}. 
\begin{deluxetable}{llll}
\tablecaption{Parameter ranges.\label{tab:paramrange}}
\tablecolumns{4}
\tablewidth{0pt}
\tablehead{
\colhead{Parameter} &
\colhead{Range} &
\colhead{Course Step} &
\colhead{Fine Step} 
}
\startdata
$q_0$ & $1.0$ - $100.0$\tablenotemark{a} & 10\tablenotemark{a} & 5\tablenotemark{a}\\
$\alpha$ & $-3.5$ - $-1.0$ & 0.5 & 0.1\\
$\beta$ & $0.0$ - $3.0$ & 0.5 & 0.1 \\
$\gamma$ & $-3.0$ - $0.0$ & 0.5 & 0.5 \\
\enddata
\tablenotetext{a}{units of $\sn{1}{-4}$ erg~cm$^{-3}$~s$^{-1}$}
\end{deluxetable}

\subsection{Comparisons to EUNIS and EIS observations} \label{sec:compare}
The synthetic line intensities derived for each simulation are integrated along an Earth observer line of sight and projected onto the EIS and EUNIS image maps obtained in \S\ref{sec:obs} using the following technique. The 3D volume over which our simulations were performed, as described in \S\ref{sec:nanoflares}, is divided into many voxels chosen to be sufficiently small so that their volumes project to within a single EUNIS or EIS pixel. The heliocentric Cartesian coordinates of these voxels are projected onto the EUNIS and EIS pixel grids using the SSW IDL routines, \proc{WCS\_CONV\_HCC\_HPC} and \proc{WCS\_GET\_PIXEL}. A simulated image is obtained by summing the contribution to each pixel from each voxel, and finally the simulated image is convolved with the EUNIS and EIS point spread functions. As an example, Figure~\ref{fig:comparelines} compares EIS and EUNIS images of several lines with the corresponding simulated images. 

The end result of each simulation is a set of images for the lines listed in Table~\ref{tab:linelist}. We compare these with the EUNIS and EIS observations %. First, we decrease the spatial resolution of the simulated and observed images to $4 \times 4$ arcsec. This is to ensure that any small misalignment between images does not greatly affect the final result. Then we calculate
by calculating the weighted L2-norm difference, $d$, between the simulated and observed images. That is 

\begin{equation}
 d = \frac{1}{N_i N_l} \sqrt{\sum\limits_l \sum\limits_i \frac{\left(I_{li}^{obs} - I_{li}^{sim}\right)^2}{(E_{li}^{obs})^2}}
\end{equation}
where $I_{li}^{sim}$, $I_{li}^{obs}$, and  $E_{li}^{obs}$ are the simulated, observed and uncertainty in observed intensities in the $i^{th}$ pixel of the $l^{th}$ line, and $N_i$ and $N_l$ are the number of pixels and lines, respectively. Our parameter study seeks to minimize $d$. 

\subsection{Results and Discussion} \label{sec:results}
\begin{figure*}
 \epsscale{.9}
 \plotone{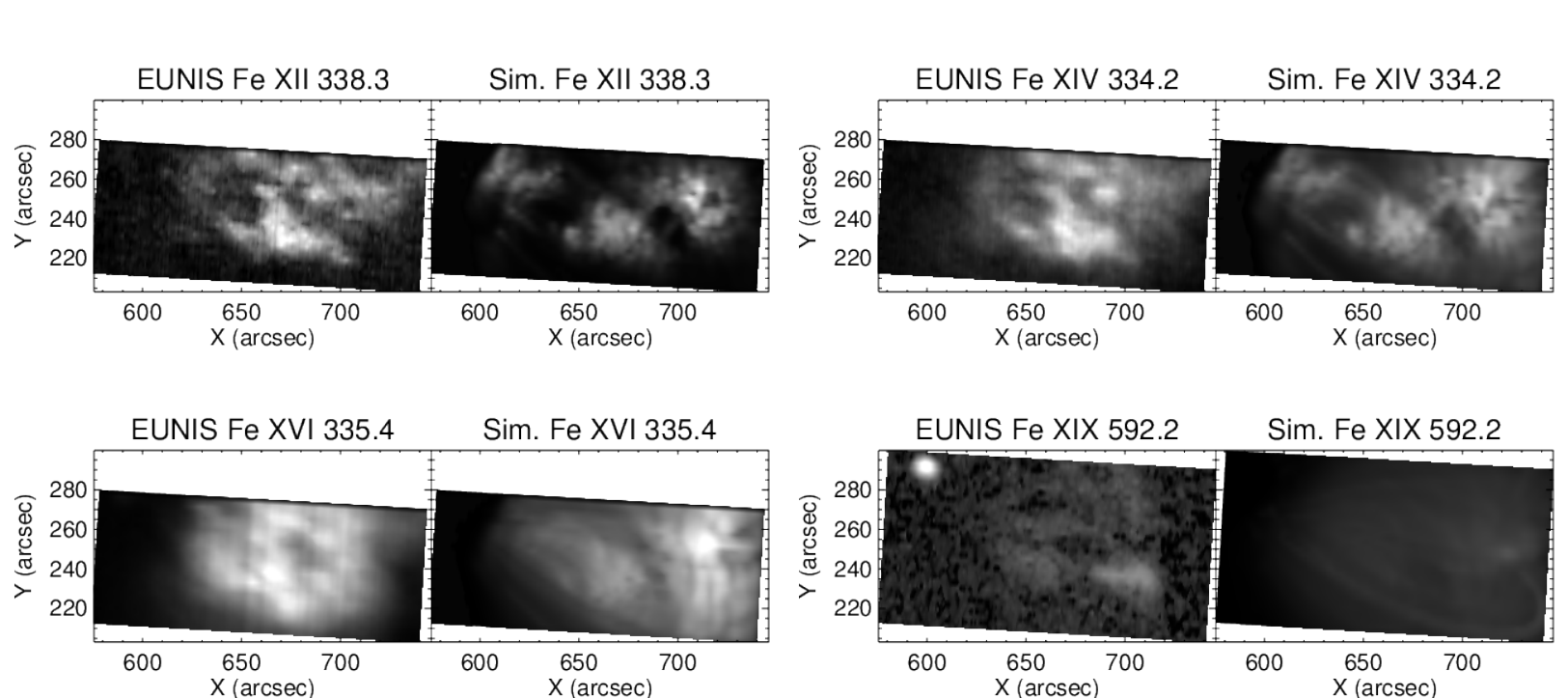}
 \plotone{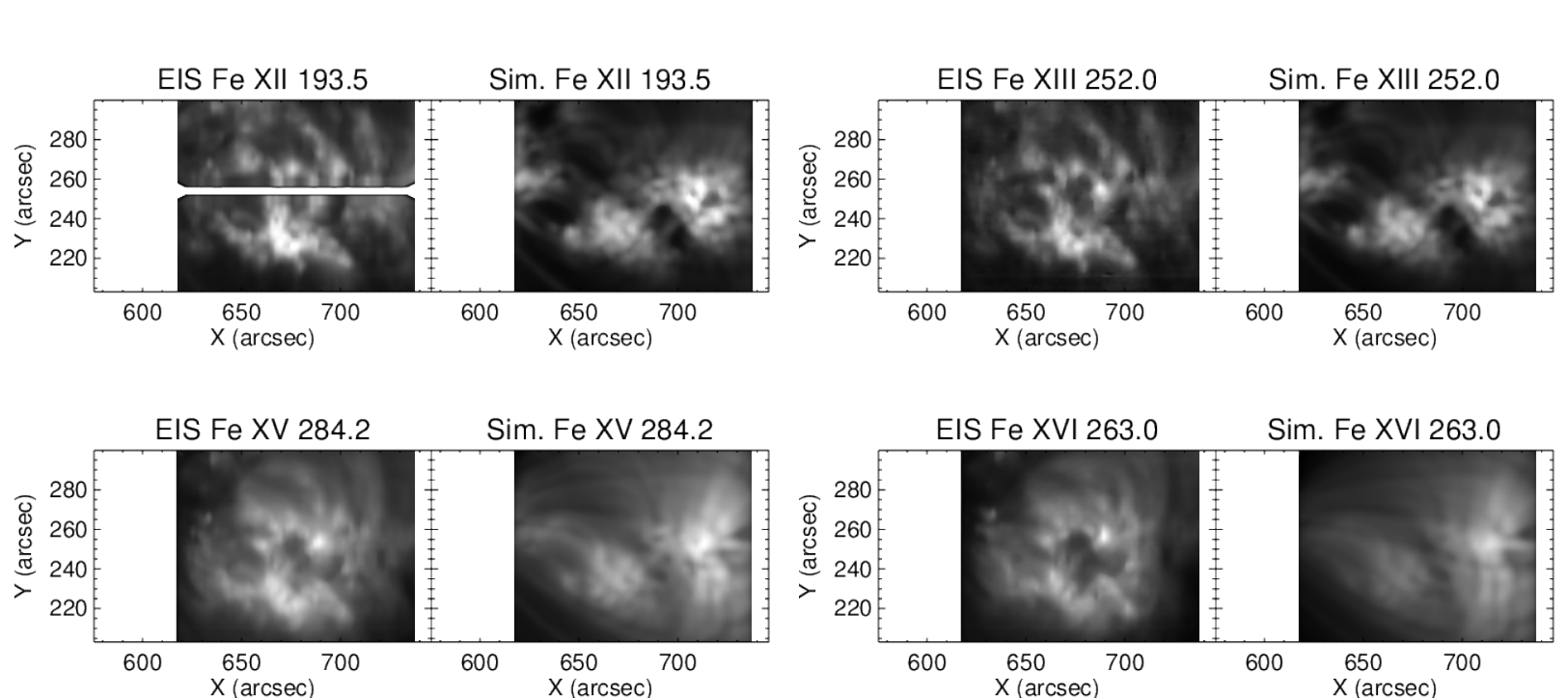}
 \caption{Comparisons of several line observed by EUNIS (top two rows) and EIS (bottom two rows) to synthetic images from the best fit parameter set $q_0 = \sn{1.0}{-3}$ erg cm$^{-3}$ s$^{-1}$, $\alpha = -2.4$, $\beta = 1.5$, and $\gamma = -1.0$. \label{fig:comparelines}}.
\end{figure*}

\begin{figure*}
  \epsscale{.9}
  \plotone{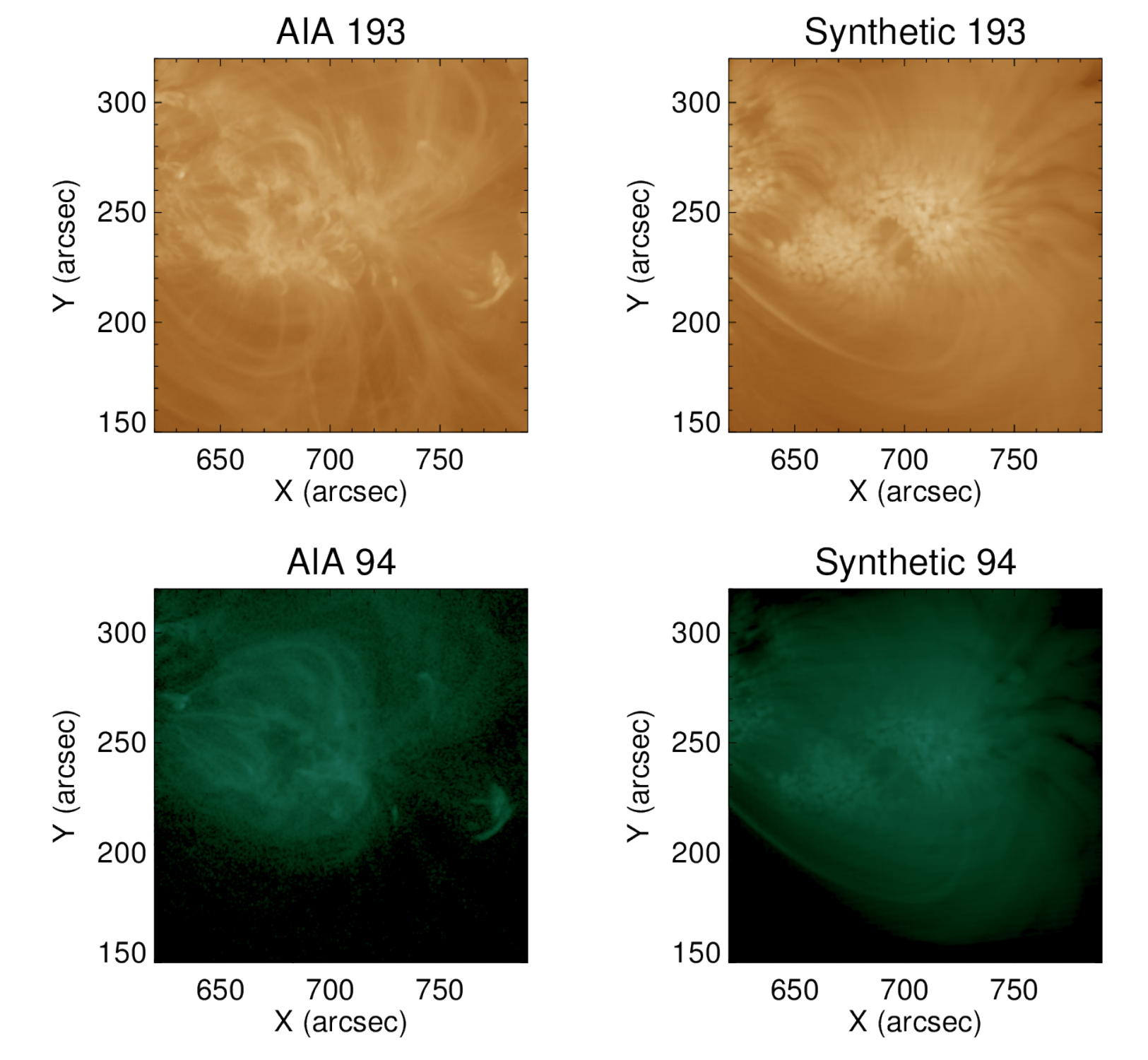}
 \caption{Comparisons of the observed AIA 193 and 94 channels to synthetic images produced using the best fit parameter set. \label{fig:compareaia}}.
\end{figure*}

We performed approximately 500 simulations varying $(q_0, \alpha, \beta, \gamma)$ over the ranges in Table~\ref{tab:paramrange}. Since each simulation requires evolving more than 2800 loops, this represents a major computational undertaking. 
%In Figure~\ref{fig:dscatter}, we show a scatter plot of the values we obtained for $d$. We have labeled several of these using the notation, ($q_0$, $\alpha$, $\beta$, $\gamma$). 
We found the parameter set that minimizes $d$ to be $q_0 = \sn{1.0}{-3}$ erg cm$^{-3}$ s$^{-1}$, $\alpha = -2.4$, $\beta = 1.5$, and $\gamma = -1.0$. Figure~\ref{fig:comparelines} compares the predicted emission in several lines with the EIS and EUNIS images. Similarly, Figure~\ref{fig:compareaia} compares AIA 193 and 94 channel images with synthetic images constructed from our best fit parameter set.

\begin{figure}
 \plotone{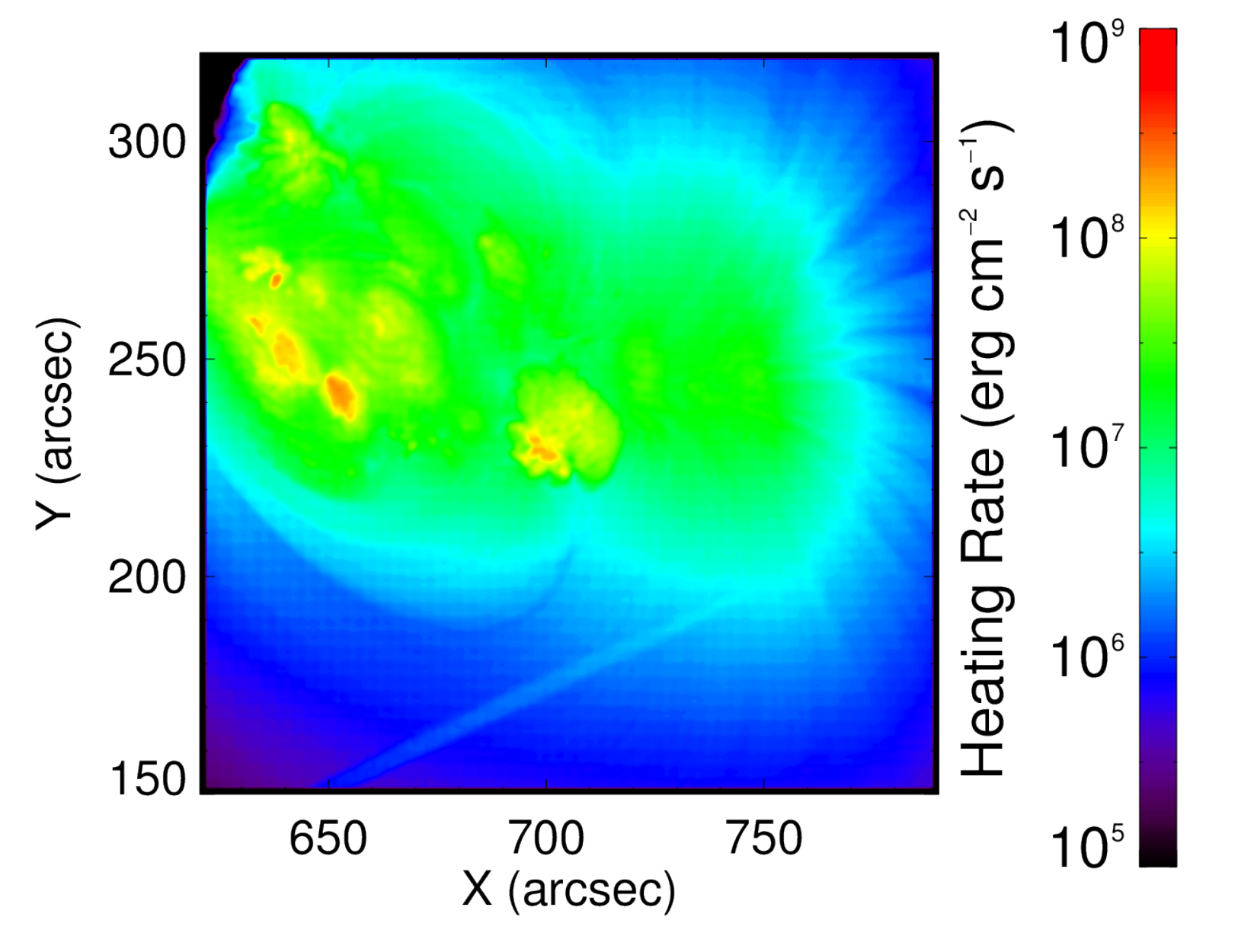}
 \caption{The volumetric heating rate resulting from our best fit simulation integrated along the line of sight.\label{fig:hr}}
\end{figure}

\begin{figure*}
 \plotone{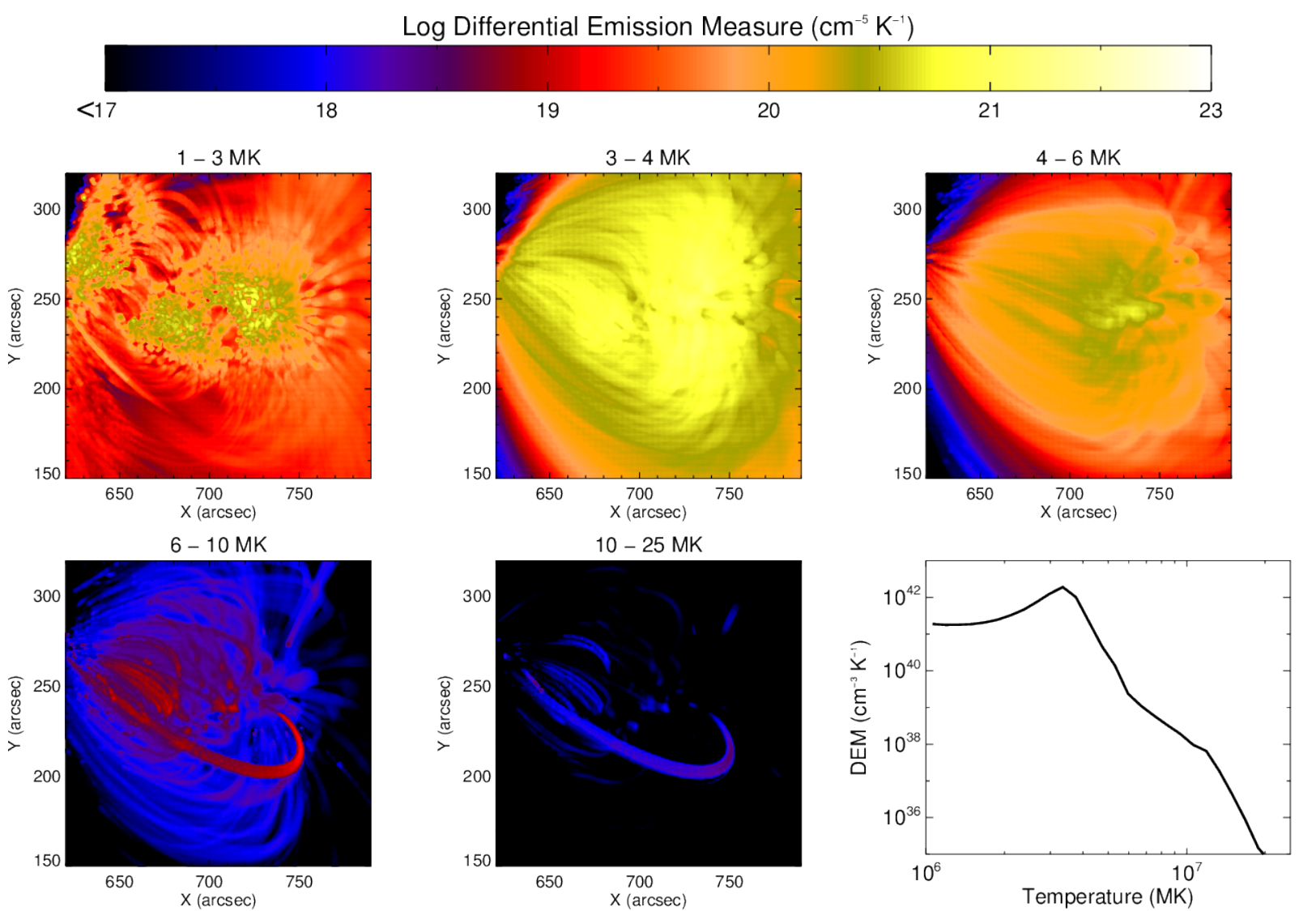}
 \caption{(Top row and left two panels of bottom row) The DEM in several temperature bins produced from the simulation with the best match parameter set. (Bottom right) The spatially integrated DEM. \label{fig:dempanels}}
\end{figure*}

Figure~\ref{fig:hr} shows the time-averaged heating rate (Eq.~\ref{eq:hr}) resulting from this simulation integrated along the line of sight. Most of the active region receives $\sim 10^7$ erg cm$^{-2}$ s$^{-1}$ as expected from previous studies \citep[e.g.][]{1977ARA&A..15..363W}. However, the hottest regions receive nearly two orders of magnitude more heat. Figure~\ref{fig:dempanels} shows the time-averaged DEM resulting from this parameter set for five temperature bins. The bottom right panel shows this DEM integrated over its spatial extent. The DEM peaks at a temperature of 3.3 MK and quickly falls off, with an average slope of $-9$ in log-log space. The DEM at 10 MK is nearly 5 orders of magnitude smaller than at the peak at 3.3 MK. These values are comparable to those obtained by \citet{2012ApJ...759..141W} who studied hot emission in several active regions using the EIS instrument. By directly imaging thermal soft X-ray bremsstrahlung from AR~12234 during the FOXSI-2 sounding rocket flight \citep{doi:10.1117/12.2232262}, \citet{2017NatAs...1..771I} found a steeper slope of $-12$.
%Interestingly, although our predicted DEM is very small in the range of 10 - 20 MK, the thermal bremsstrahlung emission from 5 - 7 keV soft X-rays should be visible to a FOXSI like instrument given an integration time $\ge$ 1000~s. To demonstrate this we plot the spatially integrated X-ray spectrum predicted from this simulation (obtained using the SSW IDL routine \proc{F\_VTH}) in Figure~\ref{fig:xrayspec}. The 50\% contribution to the emission as a function of temperature is over plotted.

It is useful to compare our best fit parameters with predictions from coronal heating theories. Table~5 in \citet{2000ApJ...530..999M} and extended in Table~1 of \citet{2008ApJ...689.1388L} provide concise summaries for how many of these models scale with $B$, $L$, $R$, and $v$, where $R$ is the loop radius and $v$ is the footpoint velocity. Our assumption, that the loop radius scales to conserve magnetic flux means that $R \propto B^{-1/2}$. For this work, we have not investigated the scaling with velocity, so to compare with these listed models we must make an assumption of how $v$ scales with $B$ and $L$. We follow the Case ($b$) assumption from \citet{2008ApJ...689.1388L} that is $v \propto B^{-1/2}$ which is relevant for twisting-type velocities. In our best fit model, the heating rate scales as $B^{1.5}$ and $L^{-1}$. This is most similar to the scaling obtained from the ``critical angle'' model \citep{1988ApJ...330..474P,1993PhRvL..70..705B} i.e., Model~2 in \citet{2000ApJ...530..999M}. This model postulates that coronal heating occurs by dissipating magnetic energy when misalignment between adjacent flux tubes reaches a critical angle. This misalignment can result from either twisted or braided field lines. 

\section{Conclusions} \label{sec:conc}
Using the EUNIS and EIS observations of AR~11726, we have constructed images of the intensity in many atomic lines that form over a wide range of temperatures. We have used HMI magnetograms and AIA images to construct a 3D model of the magnetic field within a volume that encloses AR~11726. We have traced magnetic loops within this volume and performed simulations of how their temperature and density stratification respond to nanoflare heating. We have parameterized the nanoflare frequency distribution (Eq.~\ref{eq:freq}) and by varying these parameters and comparing predicted emission to the EUNIS and EIS observations we have constrained the frequency distribution. We find the distribution is best fit with $q_0 = \sn{1.0}{-3}$ erg cm$^{-3}$ s$^{-1}$, $\alpha = -2.4$, $\beta = 1.5$, and $\gamma = -1.0$. This scaling with magnetic field and loop length best matches the ``critical angle'' model \citep{1988ApJ...330..474P,1993PhRvL..70..705B}. 

\acknowledgments
EUNIS was supported by the NASA Heliophysics Division through its Low Cost Access to Space program. J.C.A and A.N.D acknowledge funding support through NASA's Science Innovation Fund. 

\bibliography{ar11726paper}

\end{document}